*Costantino Sigismondi*

# Measuring Danjon index and umbral magnitude of a partial lunar eclipse (July 16, 2019)

Costantino Sigismondi *(ICRA/Sapienza and ITIS Ferraris, Roma)*
submitted August, 16 2019                    *sigismondi@icra.it*

**Abstract**

The partial lunar eclipse of July 16, 2019, left the lower part of the Moon illuminated at its maximum phase in Padova (Italy). Occulting it behind far buildings it was possible to compare the light of Jupiter and Saturn defocused to the same diameter of the Moon with the light from the umbra.
The luminosity of the eclipsed Moon as well as the Danjon index have been estimated and compared with ephemerides. January 21, 2019, July 27, 2019 and September 28, 2015 total lunar eclipses data are also published.

**Sommario**

L'eclissi parziale di Luna del 16 luglio 2019 ha lasciato la parte inferiore della Luna illuminata, vista da Padova. Occultando questa parte dietro degli edifici distanti, è stato possibile valutare la magnitudine luminosa della regione in ombra della Luna comparandola con Giove e Saturno sfuocati fino ad un diametro pari a quello lunare. Anche l'indice di Danjon è stato valutato e comparato con quello previsto dalle effemeridi

**Introduction**

André Danjon former director of Paris Observatory studied the correlation between the appearances of the total lunar eclipses and the solar activity. During maximum solar activity the corona is larger and casts more light on the umbral region of the eclipse, mainly through the upper Earth's atmosphere. The presence of aerosols in the stratosphere changes the Danjon index and the visual magnitude of the eclipsed Moon (figure, Keen, 2016). Here, firstly, a method for partial eclipses.

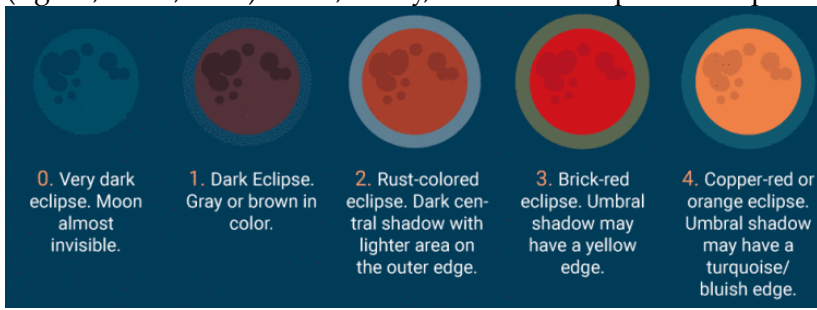

**Method for the observations**

The partial eclipse of July 16 left the lower part of the Moon illuminated, and choosing a position with buildings next to it the usual comparison between the Moon and nearby stars (Espenak, 2014) has been done.

At 23:40 local time in Padova (21:40 UT) of July 16, 2019, location Ponti Romani, Saturn<Eclipsed Moon (Umbra only)<<Jupiter. Applying the Argelander method as in variable stars observations (Yendell, 1905) in a scale of 5 Saturn was 0, the Moon was 1 and Jupiter 5. From calsky.com ephemerides Jupiter -2.5 and Saturn 0.1 magnitude.

Then the umbral part of the Moon was 0.1-[2.6/5]=-0.42 mag.

An errorbar of 0.1 magnitudes is appropriate.

The color of the umbral zone was brown dark with silver borders, so with Danjon index 1.

To compare with other total eclipses we have to rescale the magnitude of the umbral zone to the whole disk.

Being the geometrical magnitude of the eclipse 0.66, the shadowed area is ≤ 81%. The scaling factor is x 1.2345, that for the Pogson law becomes 0.23 magnitudes. So the rescaled magnitude for a whole eclipsed disk is -0.65±0.10 mag.

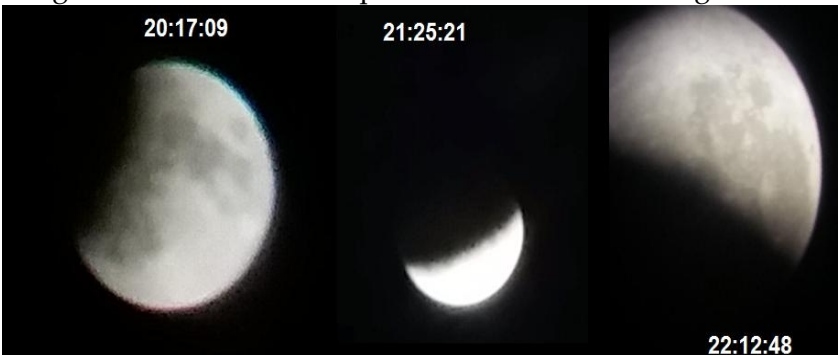

Fig. 2 Collage of 16 July 2019 eclipse photo with UT timings

Costantino Sigismondi

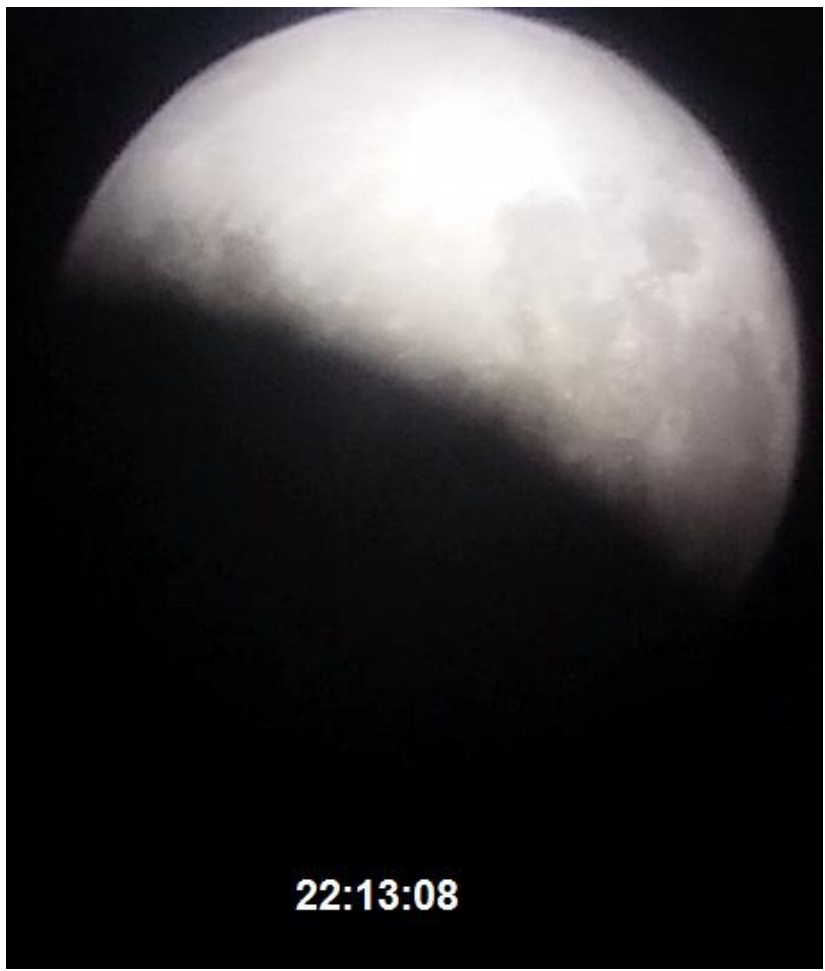

**Comparison with calsky.com ephemerides**
The Danjon index is calculated only for total eclipses and the brigthness of the eclipsed Moon includes the illuminated part*

| Date | Geom. Mag. | Visual Mag. | Danjon index |
|---|---|---|---|
| 28 SEP 15 | 1,282 | -1,3 (-1 oss.) | 2,3 (2 oss.) |
| 27 JUL 18 | 1,614 | 0,6 (0.5 oss.) | 1,3 (1,5 oss.) |

| 21 JAN 19 | 1,201 | -1,9 (-1.95±0.05) | 2,8 (3 oss.**) |
|---|---|---|---|
| 16 JUL 19 | 0,658 | -9,0*(-0,65 oss) | / (1 oss.) |
| 22 MAY 22 | 1,420 | -0,3 | 1,7 |
| 7 SET 25 | 1,368 | -0,6 | 1,9 |
| 31 DEC 28 | 1,232 | -1,6 | 2,5 |
| 26 JUN 29 | 1,850 | 1,5 | 0,9 |
| 20 DEC 29 | 1,122 | -2,4 | 3,3 |
| 18 OTT 32 | 1,109 | -2,5 | 3,5 |
| 14 APR 33 | 1,098 | -2,7 | 3,9 |
| 11 FEB 36 | 1,305 | -1,1 | 2,2 |

The observed Danjon index of July 27 2018 eclipse (Sigismondi, 2018) was in agreement with the ephemerides, as well as the visual magnitude. The eclipse of Jan 21, 2019 was clouded out on most central Italy, so I don' t have direct observations. If the Danjon index is linked to the solar cycle (e.g. Matsushima, 1966), the present eclipse should be similar appearances to the eclipses around 2030, because of the solar cycle, or 2008 going back in time. The eclipses around 2030 have both luminosities at maximum eclipse around -2.5 and Danjon indexes around 3.4, making them rather bright total lunar eclipses (as the 21 Jan 2019 one). The eclipse of 2018 is near the solar minimm after cycle 24 and it is dark, while the 2019 Jan 21 is not so dark. Aperture photometry on the photo made by Daniel Bottcher (**) at Arlington Heigths Chicago at 22:15 UT of Jan 21, 2019 fournished the magnitudes results for that eclipse, his decription gave the Danjon index.

**Conclusions**

The periodicity of 11 years on the Danjon index is not evident from the ephemerides, because of the influence of the geometrical magnitude of the eclipse; so an eclipse with the lunar lim grazing the Earth's umbral limit appears brighter than another one with the center of the Moon close to the umbral center, like on June 26, 2029.

The results of the partial eclipse of July 16, 2019, reported in term of total eclipse by scaling the area with respect to the visual magnitude suggest a classification of Danjon index class 1 and visual magnitude -0.65±0.10.

The solar activity is at its deepest minimum between cycles 24 and 26.

The Danjon index is also related with the presence of aerosols in the stratosphere, and its monitoring can be an interesting task for knowing the conditions of our atmosphere and the climatic implications for our planet.

**Acknowledgments** To Daniel Bottcher, Arlington Heights School, Illinois USA for the photo of the Eclipse of January 21, 2019, sent with his subjective descriptions (for Danjon index).

**References**

P. S., Yendell, Journal: Popular Astronomy, 13, 453 (1905).

S. Matsushima, Astronomical Journal, 71, 699 (1966)

R. Keen (2016)
https://www.esrl.noaa.gov/gmd/publications/annual_meetings/2016/posters/P60-Keen.pdf

F. Espenak, http://www.mreclipse.com/Special/danjon.html (2014)

F. Espenak (2009)
https://eclipse.gsfc.nasa.gov/LEplot/LEplot2001/LE2019Jul16P.pdf

C. Sigismondi, http://www.icra.it/gerbertus/2018/Gerb-11-2018-Sigismondi-penumbra-27-32.pdf (2018)

C. Sigismondi, http://www.icra.it/gerbertus/2016/Gerb-9-2016-Sigismondi-eclisse-penombra-91-94.pdf (2016)

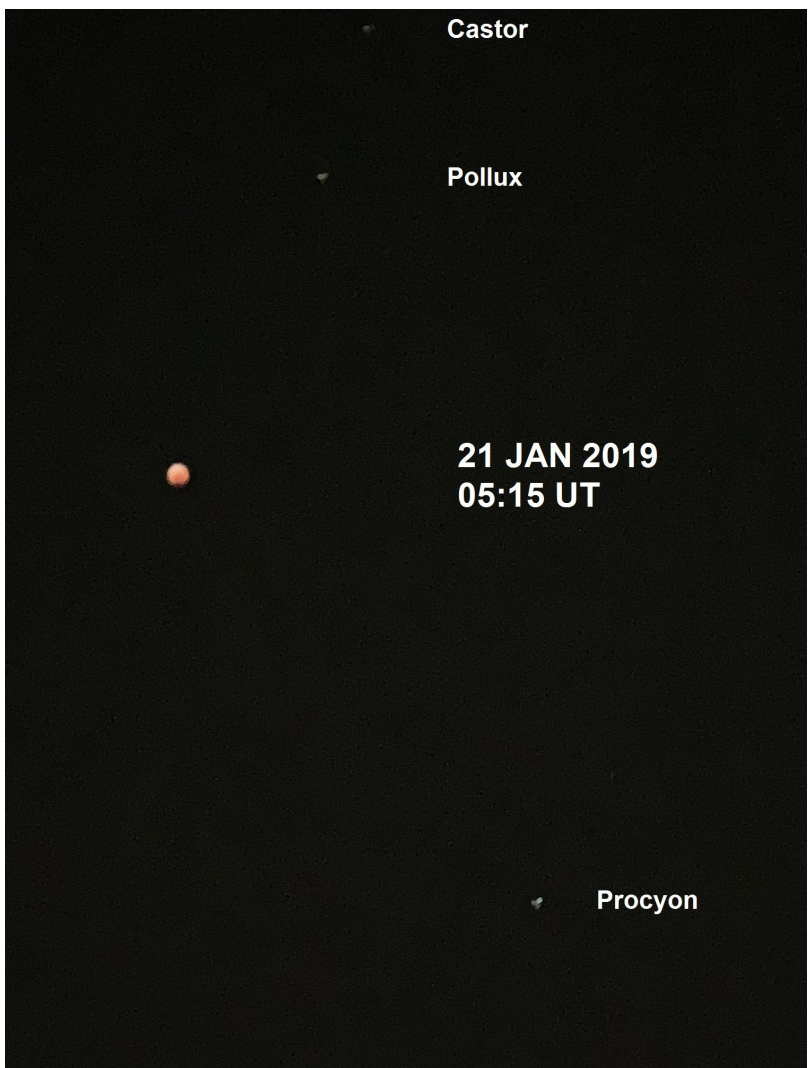

Photo of the eclipse of 21 Jan 2019 at its maximum phase.
The comparison stars are labeled, the image is moved, it avoided saturated points in the digital image, making more precise the aperture photometry. The full eclisped Moon is about 3.2 magnitudes brighter than Procyon and 2.7 than Pollux: and their different altitudes have been considered to correct magnitudes for airmasses.